\documentclass[aps,showkeys,preprintnumbers,amsmath,amssymb,nofootinbib]{revtex4}
\usepackage{mathrsfs}
\usepackage{graphicx}
\usepackage{amsfonts}
\usepackage{epstopdf}
\usepackage{eurosym}
\usepackage{latexsym}
\usepackage[center]{subfigure}
\providecommand{\U}[1]{\protect\rule{.1in}{.1in}}
\usepackage{color}%

 \newcommand{\bq}{\begin{equation}}
 \newcommand{\eq}{\end{equation}}
 \newcommand{\bqn}{\begin{eqnarray}}
 \newcommand{\eqn}{\end{eqnarray}}
 \newcommand{\nb}{\nonumber}
 \newcommand{\lb}{\label}
\begin{document}
\baselineskip=0.6 cm

\title{Analysis of s-wave, p-wave and d-wave holographic superconductors in Ho\v{r}ava-Lifshitz gravity}

\author{Kai Lin$^{1,2)}$}\email{lk314159@hotmail.com}
\author{Xiao-Mei Kuang$^{3,4)}$}\email{xmeikuang@gmail.com}
\author{Wei-Liang Qian$^{2,5)}$}\email{wlqian@usp.br}
\author{Qiyuan Pan$^{6)}$}\email{panqiyuan@126.com}
\author{A. B. Pavan$^{1)}$}\email{alan@unifei.edu.br}

\affiliation{1) Universidade Federal de Itajub\'a, Instituto de F\'isica e Qu\'imica, 37500-903, Itajub\'a, MG, Brazil}
\affiliation{2) Escola de Engenharia de Lorena, Universidade de S\~ao Paulo, 12602-810, Lorena, SP, Brazil}
\affiliation{3) Center for Gravitation and Cosmology, College of Physical Science and Technology, Yangzhou University, Yangzhou 225009, China}
\affiliation{4) Instituto de F\'isica, Pontificia Universidad Cat\'olica de Valpara\'iso, Casilla 4059, Valpara\'iso, Chile}
\affiliation{5) Faculdade de Engenharia de Guaratinguet\'a, Universidade Estadual Paulista, 12516-410, Guaratinguet\'a, SP, Brazil}
\affiliation{6) Department of Physics, Key Laboratory of Low Dimensional Quantum Structures and Quantum Control of Ministry of Education, and Synergetic Innovation Center for Quantum Effects and Applications, Hunan Normal University, Changsha, Hunan 410081, China.}

\date{\today }

\begin{abstract}

In this work, the s-wave, p-wave and d-wave holographic superconductors in the Ho\v{r}ava-Lifshitz gravity are investigated in the probe limit.
For the present approach, it is shown that the equations of motion for different wave states in Einstein gravity can be written into a unified form, and condensates take place in all three cases.
This scheme is then generalized to Ho\v{r}ava-Lifshitz gravity, and an unified equation for multiple holographic states is obtained.
Furthermore, the properties of the condensation and the optical conductivity are studied numerically.
It is found that, in the case of Ho\v{r}ava-Lifshitz gravity, it is always possible to find some particular parameters in the corresponding Einstein case where the condensation curves are identical. For fixed scalar field mass $m$, a non-vanishing $\alpha$ becomes the condensation easier than in Einstein gravity for s-wave superconductor. However, the p-wave and d-wave superconductors have $T_c$ greater than s-wave one.
\end{abstract}


\maketitle
\newpage
\vspace*{0.2cm}

\section{Introduction}
\renewcommand{\theequation}{1.\arabic{equation}} \setcounter{equation}{0}

The black hole is one of the most intriguing subjects in physics.
The discovery of thermal Hawking radiation and subsequent black-hole thermodynamics have led to a profound impact on the understanding of quantum gravity, by unifying the theories of gravity, quantum mechanics, and statistical physics.
Moreover, the anti-de Sitter/conformal field theory (AdS/CFT) correspondence establishes a relationship between the theory of quantum gravity in $n+1$ dimensional AdS spacetime and the conformal field theory on the $n$ dimensional boundary\cite{Maldacena,Gubser,Witten}.
The strong-weak nature of the conjectured duality indicates when the quanta of the conformal field theory are strongly coupled, those in the gravitational theory are weakly interacting, and thus become more accessible mathematically.
Therefore, the AdS/CFT correspondence provides us a potent tool to investigate a strongly interacting system via the study of its classical dual theory with weak coupling.
In particular, the dual gravity system can be approximated by general relativity where the notion of temperature is achieved by that of the Hawking radiation of the black hole in the bulk.

Following Maldacena's pioneering insight in 1997 \cite{Maldacena2}, many different implementations of the AdS/CFT correspondence have been discovered.
The study of the s-wave superconductor was first realized \cite{Hartnoll:2008vx,Hartnoll:2008kx} by introducing a complex scalar field coupled to a $U(1)$ gauge field in an asymptotically AdS black-hole background.
It is justified as one tunes the temperature and/or chemical potential, the gravitational system might become unstable.
Eventually, this leads to the emergence of additional ``hair" in the bulk, when the scalar field acquires a nonvanishing vacuum expectation value.
The latter is translated into the condensation of a composite charged operator in the dual lower dimensional field theory, which corresponds to the superconducting phase transition in question.
As the phase transition takes place, the global $U(1)$ symmetry on the boundary is spontaneously broken as described phenomenologically by the Ginzburg-Landau theory.
It was shown that this simple holographic setup indeed captures the key features of condensation and optical conductivity observed in real superconductors.

Subsequently, many holographic superconductors were proposed by providing various solvable models of strongly coupling high-temperature superconductivity.
The s-wave holographic superconductors were extended and investigated by many authors
\cite{Albash:2008eh,Herzog:2008he,Albash:2009iq,Gregory:2009fj,Konoplya:2009hv,Franco:2009if,Chen:2009vz,Pan:2009xa,
Ge:2010aa,Momeni:2010jf,Liu:2010ka,Myung:2010rb,Liu:2010bq,Kuang:2010jc,Barclay:2010up,Wu:2010vr,Flauger:2010tv,
Cai:2011ky,Peng:2011gh,Li:2013fhw,Kuang:2013oqa,Roychowdhury:2012hp,Fan:2013tga,Dey:2014xxa,Ling:2014laa,Lu:2013tza,Kuang:2016edj}.
Moreover, studies have further generalized the model to describe superconductors with different wave states.
This is genuinely interesting since the superconductivity in high-temperature cuprates is well-known to be d-wave type \cite{Tsuei}.
The p-wave and d-wave superconductor models can be constructed by introducing vector and spin two fields in the place of the scalar field.
The models of p-wave superconductor mainly include the SU(2) Yang-Mills p-wave model \cite{Gubser:2008wv}, the Maxwell-vector p-wave model \cite{Cai:2013pda} and the helical p-wave model \cite{Donos:2011ff}.
On the other hand, the d-wave superconductor can be implemented by the CKMWY d-wave model \cite{Chen:2010mk} proposed by Chen et al.
There, the authors tackled the problem with a minimal action consists of a symmetric, traceless second-rank tensor field and the Maxwell field in the background of the AdS black-hole metric.
The system is found to be superconducting in the DC limit but has no hard gap.
By analyzing the coupling terms of the action in more details, the d-wave model was also extensively investigated by other authors \cite{Benini:2010pr,Zeng:2010vp,Ge:2012vp,Kim:2013oba,Nishida:2014lta,Li:2014wca,Krikun:2015tga}.
For reviews on holographic superconductors, see for example \cite{Horowitz:2010gk,Cai:2015cya} and the references therein.

Owing to the difficulties in renormalization of general relativity, Ho\v{r}ava \cite{HL} proposed Ho\v{r}ava-Lifshitz theory to cope with the problem by explicitly breaking the Lorentz invariance.
In this theory, space and time are no longer symmetric and satisfy the following foliation-preserving diffeomorphism,
\bqn
 \lb{0}
t'=f(t),~~~~~~x'^i=\zeta^i(i,x^k), (i,k=1,2,3)
\eqn
In order to be consistent with the experimental observations, the breaking of Lorentz invariance is negligible in the lower energy region where the theory restores general relativity.
After Ho\v{r}ava-Lifshitz theory was proposed, it encountered various challenges.
By introducing a scalar $U(1)$ gauge field and weakly broken detailed balance condition, problems such as ghost mode, instability, and strong coupling have been solved \cite{HW1}.
The redundant higher order terms in the action were also successfully removed \cite{HW2,HW3,HW4}, and therefore the prediction power of the theory was improved.
By calculating the classical limit of the model, the post-Newtonian parameters were obtained in accordance with all the existing experimental data in the solar system \cite{HW3}.
Therefore, it is interesting to explore the properties of the dual holographic superconductors, particularly that of the d-wave state.

The present study involves an attempt to study the d-wave holographic superconductor for modified gravity following the CKMWY model.
In our work, we assume minimal coupling between the massive spin-two field and the Maxwell field.
It is shown that the equations of motion for different wave states can be written into an unified form, and condensates are found to take place in all three cases.
The above scheme is then generalized to the Ho\v{r}ava-Lifshitz theory.

The paper is structured as follows.
In the next section, we review the s-wave, p-wave, and d-wave holographic superconductors in the Einstein gravity.
The Maxwell-vector model is adopted to describe the p-wave state, while a massive spin-two field with minimal coupling is utilized to model the condensate in the d-wave superconductor.
In our approach, it is shown that the equations of motion can be rewritten into an unique form by a particular choice of Ansatz for the perturbative field and by properly redefinition of the matter fields.
In section III, the model is generalized to investigate the dual superconductor in Ho\v{r}ava-Lifshitz theory.
The condensation curve and the optical conductivity are calculated numerically. A semi-analytical result for the condensation curve and critical temperature are obtained and discussed.
The qualitative behavior of the dual superconductor is discussed.
The concluding remarks are given in the last section.

\section{Holographic superconductors of different wave states in Einstein Gravity}\label{sec:HSinEinstein}
\renewcommand{\theequation}{2.\arabic{equation}} \setcounter{equation}{0}

In this section, we first study the holographic setups of s-wave, p-wave and d-wave superconductors in Einstein gravity, which has been widely investigated in the literatures.
Furthermore, it is shown that the equations of motion can be rewritten into an unified form by an appropriate choice of Ansatz and redefinition of the fields.
It is noted that the $r$ component of Maxwell's equations usually implies that the phase of the matter field must be constant.
Therefore, without loss of generality, we take the matter field to be real from now on.
Also, in this work, we only consider the probe limit and ignore the backreaction on the gravitational field.

For the s-wave holographic superconductor, the bulk action of the matter fields proposed in \cite{Hartnoll:2008vx} can be written as
\bqn\lb{2}
S_S=\frac{1}{16\pi G}\int d^4x\sqrt{-g}\left[{\cal L}_F+{\cal L}_S+{\cal L}_{C_S}\right]
 \eqn
with
 \bqn
 \lb{3}
{\cal L}_F&=&-\frac{1}{4}F_{\mu\nu}F^{\mu\nu},~~
{\cal L}_S=-\left(\partial_\mu\Psi\right)\left(\partial^\mu\Psi\right),~~
{\cal L}_{C_S}=-\left(q^2A_\nu A^\nu+m_S^2\right)\Psi^2.
 \eqn
where $F_{\mu\nu}=\nabla_\mu A_\nu-\nabla_\nu A_\mu$ is the strength of Maxwell field and $\Psi$ is the scalar field with the mass $m_S$.
We consider a four dimensional planar black hole background as
 \bqn
 \lb{4}
ds^2=-f(r)dt^2+\frac{dr^2}{f(r)}+r^2(dx^2+dy^2).
 \eqn
We take the Ansatz for the matter fields, $\Psi=\psi_S(r)$ and $A_\nu=\delta^t_\nu \phi(r)$.
By using the variational principle, the action gives the following equations of motion
\begin{subequations}\lb{5}
 \bqn\lb{5a}\psi_S''+\left(\frac{f'}{f}+\frac{2}{r}\right)\psi_S'+\left(\frac{q^2\phi^2}{f^2}-\frac{m_S^2}{f}\right)\psi_S&=&0,\eqn
 \bqn\lb{5b}\phi''+\frac{2}{r}\phi'-\frac{2q^2\psi_S^2}{f}\phi&=&0 .\eqn
\end{subequations}

In order to study the conductivity, we consider the perturbation of the Maxwell field $ A_\nu=\delta_\nu^xA_x(r)\exp(-i\omega t)$ and obtain
\bqn\lb{6}
 A_x''+\frac{f'}{f}A_x'+\left(\frac{\omega^2}{f^2}-\frac{2q^2\psi_S^2}{f}\right)A_x = 0.
\eqn

By solving the above Eqs.(\ref{5a}) and (\ref{5b}), one can study the phase transition of the s-wave holographic superconductor through the condensation of the scalar field.
The coupled Eqs.(\ref{5}-\ref{6}) determine the correlation function of s-wave holographic superconductors.
The perturbation equation of the Maxwell field, Eq.(\ref{6}), can be evaluated using the obtained solution of Eqs.(\ref{5a}) and (\ref{5b}), and is used to explore the conductivity of the superconducting state.

To study the p-wave holographic superconductor, various approaches have been proposed.
Among others, they consist of $SU(2)$ Yang-Mills p-wave model \cite{Gubser:2008wv} with non-Abelian vector field, Maxwell-vector p-wave model \cite{Cai:2013pda} and the helical p-wave model \cite{Donos:2011ff}.
In this paper, we make use of the Maxwell-vector p-wave model with the following action
 \bqn
 \lb{7}
S_P&=&\frac{1}{16\pi G}\int
d^4x\sqrt{-g}\left[-\frac{1}{4}F_{\mu\nu}F^{\mu\nu}-m_P^2\rho^\mu\rho_\mu+iq\gamma\rho_\mu\rho_\nu
F^{\mu\nu}\right.\nb\\
&&\left.-\frac{1}{2}\left(\nabla^\mu\rho^\nu-\nabla^\nu\rho^\mu
+iqA^\mu\rho^\nu-iqA^\nu\rho^\mu\right)\left(\nabla_\mu\rho_\nu-\nabla_\nu\rho_\mu
-iqA_\mu\rho_\nu+iqA_\nu\rho_\mu\right)\right]\nb\\&=&\frac{1}{16\pi G}\int d^4x\sqrt{-g}\left[{\cal L}_F+{\cal
L}_P+{\cal L}_{C_P}\right] ,
 \eqn
where ${\cal L}_F$ is defined above in Eq.(\ref{3}) and
 \bqn
 \lb{8}
{\cal
L}_P&=&-\frac{1}{2}\left(\nabla^\mu\rho^\nu-\nabla^\nu\rho^\mu\right)\left(\nabla_\mu\rho_\nu-\nabla_\nu\rho_\mu\right),~~
{\cal L}_{C_P}=-\left(q^2A_\nu
A^\nu+m_P^2\right)\rho^\mu\rho_\mu+iq\gamma\rho_\mu\rho_\nu
F^{\mu\nu} ,
 \eqn
where $m_P$ is the mass of the vector field $\rho^\mu$.
By taking $\rho_\mu=\delta_\mu^x\rho_x(r)$ and redefining $\rho_x=r\psi_P$, the action Eq.(\ref{7}) leads to the following equations of motion for the matter fields in the background metric Eq.(\ref{4}),
\begin{subequations}\lb{9}
 \bqn\lb{9a}\psi_P''+\left(\frac{f'}{f}+\frac{2}{r}\right)\psi_P'+\left(\frac{q^2\phi^2}{f^2}-\frac{m_P^2}{f}+\frac{f'}{rf}\right)\psi_P&=&0,\eqn
 \bqn\lb{9b}\phi''+\frac{2}{r}\phi'-\frac{2q^2\psi_P^2}{f}\phi&=&0,\eqn
 \bqn\lb{9c}A_x''+\frac{f'}{f}A_x'+\left(\frac{\omega^2}{f^2}-\frac{2q^2\psi_P^2}{f}\right)A_x&=&0 .\eqn
\end{subequations}

Comparing Eqs.(\ref{9a}), (\ref{9b}) and (\ref{9c}) with Eqs.(\ref{5}-\ref{6}), one observes that Eqs.(\ref{9b}) and (\ref{9c}) are exactly the same as Eq.(\ref{5b}) and  (\ref{6}).
There is some small difference between Eq.(\ref{9a}) and Eq.(\ref{5a}), which will be further discussed below.

Now we turn to the d-wave holographic superconductor.
Following the CKMWY model \cite{Chen:2010mk}, a second rank tensor $B_{\mu\nu}$ is introduced in the background metric.
The resulting action reads
\bqn
\lb{10a}
S_D=\frac{1}{16\pi G}\int d^4x\sqrt{-g}\left[{\cal L}_F+{\cal L}_D+{\cal L}_{D_S}\right] \lb{12}
 \eqn
where
\bqn
\lb{10b}
{\cal L}_D=-\left(\nabla_\gamma B_{\mu\nu}\right)\left(\nabla^\gamma B^{\mu\nu}\right),~~~{\cal L}_{D_S}=-\left(q^2A_\mu A^\mu+m_D^2\right)B_{\mu\nu}B^{\mu\nu}
 \eqn

To obtain the equation of motion, we consider the following Ansatz for the tensor field
 \bqn
 \lb{14}
B_{\mu\nu}=(\delta_\mu^x\delta_\nu^x-\delta_\mu^y\delta_\nu^y)\frac{r^2}{\sqrt{2}}\psi_D(r).
 \eqn
By using the variational principle, the resulting equation of motion for the proposed action, Eq.(\ref{12}), in terms of $\psi_D$, $\phi$, and $A_x$, read
 \begin{subequations}\lb{15}
 \bqn \lb{15a}\psi_D''+\left(\frac{f'}{f}+\frac{2}{r}\right)\psi_D'+\left(\frac{q^2\phi^2}{f^2}-\frac{m_D^2}{f}-\frac{2}{r^2}\right)\psi_D&=&0, \eqn
 \bqn \lb{15b}\phi''+\frac{2}{r}\phi'-\frac{2q^2\psi_D^2}{f}\phi&=&0, \eqn
 \bqn \lb{15c}A_x''+\frac{f'}{f}A_x'+\left(\frac{\omega^2}{f^2}-\frac{2q^2\psi_D^2}{f}\right)A_x&=&0. \eqn
 \end{subequations}

Again, in the above equations of motions, Eq.(\ref{15b}) coincides with Eq.(\ref{5b}) for the s-wave state (or Eq.(\ref{9b}) for the p-wave mode), and Eq.(\ref{15c}) is exactly Eq.(\ref{6}) for the s-wave state (or Eq.(\ref{9c}) for the p-wave case).
The only difference exists between Eqs.(\ref{5a}), (\ref{9a}) and (\ref{15a}).

Now, it is straightforward to rewrite the Eqs.(\ref{5}-\ref{6}), (\ref{9}) and (\ref{15}) into an unified form as
\begin{subequations}\lb{16}
\bqn\lb{16a}\psi''+\left(\frac{f'}{f}+\frac{2}{r}\right)\psi'+\left(\frac{q^2\phi^2}{f^2}+V_s(r)\right)\psi&=&0, \eqn
\bqn\lb{16b}\phi''+\frac{2}{r}\phi'-\frac{2q^2\psi^2}{f}\phi&=&0,\eqn
\bqn\lb{16c}A_x''+\frac{f'}{f}A_x'+\left(\frac{\omega^2}{f^2}-\frac{2q^2\psi^2}{f}\right)A_x&=&0.\eqn
\end{subequations}
where the effective potential is
 \bqn
 \lb{17}
 V_s(r)=-\frac{m^2}{f}+s(2-s)\frac{f'}{rf}-\frac{s(s-1)}{r^2}.
 \eqn
Here, $s=0, 1, 2$ correspond to s-wave, p-wave, and d-wave holographic superconductors, respectively. From the Eqs.($\ref{16a},\ref{17}$) the BF bound can be obtained for each wave state. For a Schwarzschild-AdS$_4$ black hole, for example, the BF bound for the three wave states result in
 \bqn
 \lb{17a}
 m^2_{BF}=-9/4 + 5 s - 3 s^2.
 \eqn
Thus, the wave state of the superconductor changes this important limit in AdS/CFT correspondence. From the Eq.(\ref{17a}) we have $m^2_{BF}=-\frac{9}{4}$ for s-wave, $m^2_{BF}=-\frac{1}{4}$ for p-wave and $m^2_{BF}=-\frac{17}{4}$ for d-wave. Therefore, a p-wave superconductor allows more negative squared mass for the scalar field.

\begin{figure}[h]
\includegraphics[width=10cm]{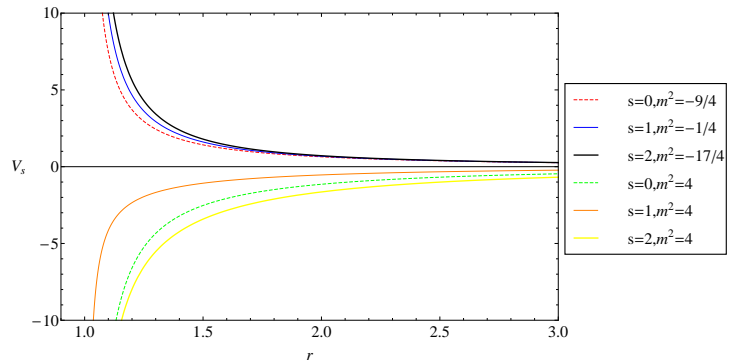}
\caption{Behaviour of $V_s$ for a Schwarzschild-AdS$_4$ with $r_h=1$ in the three different wave states. } \label{fig0}
\end{figure}
In Fig.(\ref{fig0}) one can see that around the event horizon $V_s$ is always positive for the three wave states in the BF bound and it falls faster for s-wave than the other wave states when $r$ goes to infinity. For positive values of $m$ the potential $V_s$ becomes negative for the three wave states.

\section{Holographic superconductors in Ho\v{r}ava-Lifshitz gravity}\label{sec:HSinHL}
\renewcommand{\theequation}{3.\arabic{equation}} \setcounter{equation}{0}

In this section, we generalize the study of various holographic superconductors in the previous section to the Ho\v{r}ava-Lifshitz gravity.
It is interesting to see how the breaking Lorentz invariance with $U(1)$ gauge symmetry may affect the properties of the superconductors.
We note that the Ho\v{r}ava-Lifshitz gravity has been employed to investigate the s-wave holographic superconductors in
\cite{Lin:2014bya,Luo:2016ydt}, also other applications of the AdS/CFT correspondence in condensate matter physics have been carried out \cite{Cai:2009hn,Jing:2010dc,Luo:2016zdo}.
In what follows, we extend our previous study to the p-wave and d-wave states while rewrite the equations of motion in an unified fashion.

\subsection{An unified equations of motion for holographic superconductors}

Due to the breaking of the symmetry between time and space in Ho\v{r}ava-Lifshitz gravity, it is convenient to describe the four dimensional spacetime in terms of the ADM metric\cite{Griffin:2012qx}
 \bqn
 \lb{18}
ds^2=-N^2dt^2+\gamma_{ij}\left(dx^i+N^idt\right)\left(dx^j+N^jdt\right),
 \eqn
where $i,j=1,2,3$, or equivalently, one has
 \bqn
 \lb{19}
g_{\mu\nu}&=& \left(
  \begin{array}{cc}
    g_{00} & g_{0j} \\
    g_{i0} & g_{ij} \\
  \end{array}
\right)=\left(
  \begin{array}{cc}
    -N^2+N_kN^k & N_j \\
    N_i & \gamma_{ij} \\
  \end{array}
\right),~~~
g^{\mu\nu}= \left(
  \begin{array}{cc}
    g^{00} & g^{0j} \\
    g^{i0} & g^{ij} \\
  \end{array}
\right)=\left(
  \begin{array}{cc}
    -\frac{1}{N^2} & \frac{N^j}{N^2} \\
    \frac{N^i}{N^2} & \gamma^{ij}-\frac{N^iN^j}{N^2} \\
  \end{array}
\right).
 \eqn

In order to deal with a set of difficulties in HL gravity such as, strong coupling, instability, ghost mode and redundant parameters, it is found in \cite{HW4} that one may introduce a field $A$ with $U(1)$ gauge symmetry such that the action reads
 \bqn
 \lb{20}
S&=&\frac{1}{16\pi G}\int dtd^3xN\sqrt{\gamma}\left[{\cal L}_K-{\cal L}_V+{\cal L}_A+{\cal L}_\varphi+{\cal L}_S+16\pi G{\cal L}_M\right] ,
 \eqn
where
 \bqn
 \lb{21}
{\cal L}_K&=&K_{ij}K^{ij}-\lambda K^2,~~~{\cal L}_V=2\Lambda-\beta_0a_ia^i+\gamma_1R+{\cal L}_V^H,\nb\\
{\cal L}_A&=&\frac{A}{N}(2\Lambda_g-R),~~~{\cal L}_S=\frac{A-{\cal A}}{N}\left(\sigma_1a_ia^i+\sigma_2a_i^i\right) .
\eqn
Here ${\cal L}_V^H$ represent the higher energy part of ${\cal L}_V$, and the coefficients $\lambda$, $\beta_0$, $\gamma_1$ $\sigma_1$ and $\sigma_2$ are all constants.
${\cal A}$ and ${\cal L}_\varphi$ are the functions of $\varphi$.
One can choose the gauge $\varphi=0$ to eliminate these two terms.
The term ${\cal L}_M$ is for the matter fields, which vanishes in the vacuum. Under these conditions a planar black hole solution of this HL gravity possesses the following form
 \bqn
 \lb{22}
&&N=\sqrt{f(r)},~~N^i=0,~~A=A_0\sqrt{f(r)},~~
\gamma_{ij}=\left(
  \begin{array}{ccc}
    \frac{1}{f} & 0 & 0 \\
    0  & r^2 & 0 \\
    0  & 0 & r^2 \\
  \end{array}
\right),~~f=r^2-\frac{r_h^3}{r}
 \eqn
with $\beta_0=0,c_1=c_2$. Here we have chosen $\Lambda=\Lambda_g=-3$ without loss of generality.

In what follows, we study the holographic superconductor in the background metric of Eq.(\ref{22}) in the probe limit.
Since Eq.(\ref{22}) does not consider the higher energy sector in the action, we will restrict our present study at the lower energy sector.
According to \cite{Lin:2014bya}, the Lagrangian for the matter fields read
  \bqn
 \lb{23}
\bar{{\cal L}}_F&=&-\frac{1}{4}\left\{-\frac{2}{N^2}\gamma^{ij}\left(F_{0i}-F_{ki}N^k\right)\left(F_{0j}-F_{lj}N^l\right)+\left(1+\eta_0+\eta_A\frac{A}{N}\right)F_{ij}F^{ij}\right\},\nb\\
\bar{{\cal L}}_S&=&-\left\{-\frac{1}{N^2}\left[\partial_t \Psi-N^k\partial_k \Psi\right]^2+\left(1+\alpha_0+(\alpha_A-\alpha_B)\frac{A}{N}\right)\left(\partial_i\Psi\right)\left(\partial^i\Psi\right)+\alpha_A\frac{\Psi}{N}A^k\partial_k\Psi+\alpha_0\Psi a^i\partial_i \Psi\right\},\nb\\
\bar{{\cal L}}_P&=&-\frac{1}{2}\left\{-\frac{2}{N^2}\gamma^{ij}\left(\rho_{0i}-\rho_{ki}N^k\right)\left(\rho_{0j}-\rho_{lj}N^l\right)+\left(1+\xi_0+\xi_A\frac{A}{N}\right)\rho_{ij}\rho^{ij}\right\},\nb\\
\bar{{\cal L}}_D&=&\frac{1}{N^6}\left(H_0-N^lH_l\right)^2-\frac{\gamma^{ij}}{N^4}H_iH_j-\frac{2\gamma^{lk}}{N^2}\left[\frac{1}{N^2}\left(H_{l0}-N^mH_{lm}\right)\left(H_{k0}-N^nH_{kn}\right)-\gamma^{ij}H_{li}H_{kj}\right]\nb\\
&&+2E^{ijk}D_kB_{ij}+E^{ijk}E_{ijk}+\left(1+\gamma_0+\gamma_A\frac{A}{N}\right)\left(D_kB_{ij}\right)\left(D^kB^{ij}\right),\nb\\
\bar{{\cal L}}_{C_S}&=&-\left[m_S^2+q^2(A_t-N^kA_k)^2+\left(1+\alpha_0+(\alpha_A-\alpha_B)\frac{A}{N}\right)A_iA^i\right]\Psi^2,\nb\\
\bar{{\cal L}}_{C_P}&=&-\left[m_P^2+q^2(A_t-N^kA_k)^2+\left(1+\xi_0+\xi_A\frac{A}{N}\right)A_iA^i\right]\left[-\frac{(\rho_t-N^k\rho_k)^2}{N^2}+\tau_0\rho^i\rho_i\right]\nb\\
&&+iq\gamma\left[\bar{\gamma}\frac{\left(\rho_t-N^k\rho_k\right)^2}{N^4}N^iN^j+\tilde{\gamma}\frac{\rho_t-N^k\rho_k}{N^2}\left(N^i\rho^j+N^j\rho^i\right)+\rho^i\rho^j\right]F_{ij}\nb\\
\bar{{\cal L}}_{C_D}&=&-\left[m_D^2+q^2(A_t-N^kA_k)^2+\left(1+\gamma_0+\gamma_A\frac{A}{N}\right)A_iA^i\right]\left[\frac{\iota_0}{N^4}\left(B_{00}-2B_{0k}N^k+B_{ij}N^iN^j\right)^2\right.\nb\\
&&\left.+\frac{2\iota}{N^2}\gamma^{ij}\left(B_{0i}+B_{il}N^l\right)\left(B_{0j}+B_{jk}N^k\right)+B_{ij}B^{ij}\right],
 \eqn
where
  \bqn
 \lb{23a}
H_\mu&=&\nabla_\mu B_{00}-2N^i\nabla_\mu B_{0i}+N^iN^j\nabla_\mu B_{ij}\nb\\
H_{i\mu}&=&\nabla_\mu B_{0i}-N^k\nabla_{\mu}B_{ik}\nb\\
E_{ijk}&=&\frac{1}{N}\left[K_{ki}\left(B_{0j}-N^lB_{lj}\right)+K_{kj}\left(B_{0i}-N^kB_{ki}\right)\right]\nb\\
F_{0i}&=&\partial_tA_i-\partial_i A_t\nb\\
\rho_{0i}&=&\partial_t\rho_i-\partial_i\rho_t.
 \eqn
Moreover, we removed odd power terms from the Lagrangian in order to guarantee that the resulting equations of motion are homogeneous.

The actions for the s-wave, p-wave and d-wave holographic superconductors in HL gravity are
\bqn
\lb{24a}
\bar{S}_S&=&\frac{1}{16\pi G}\int dtd^3xN\sqrt{\gamma}\left[\bar{\cal L}_F+\bar{\cal L}_S+\bar{\cal L}_{C_S}\right],
\eqn
\bqn
\lb{24b}
\bar{S}_P&=&\frac{1}{16\pi G}\int dtd^3xN\sqrt{\gamma}\left[\bar{\cal L}_F+\bar{\cal L}_P+\bar{\cal L}_{C_P}\right],
\eqn
\bqn
\lb{24c}
\bar{S}_D&=&\frac{1}{16\pi G}\int dtd^3xN\sqrt{\gamma}\left[\bar{\cal L}_F+\bar{\cal L}_D+\bar{\cal L}_{C_D}\right],
\eqn
respectively, from which the coupled equations of motion can be derived.

In order to derive the equations of motion, we adopt the same Ansatz as before, namely, $A_\nu=\delta^t_\nu \phi(r)$, $\Psi=\psi_S(r)$, $\rho_\mu=\delta_{\mu}^x\rho_x(r)=\delta_{\mu}^x r\psi_P $ and $B_{\mu\nu}=(\delta_\mu^x\delta_\nu^x-\delta_\mu^y\delta_\nu^y)\frac{r^2}{\sqrt{2}}\psi_D(r)$.
Also, to write down the equations of motion for different wave states in an unified fashion, we introduce transformations of the fields and definitions of the parameters in the actions given by the Eqs.(\ref{24a}), (\ref{24b}) and (\ref{24c}) as follows
 \begin{itemize}
  \item transformations in the action Eq.(\ref{24a}) for s-wave superconductor:
 \bqn
 \lb{28}
\phi&\rightarrow&\frac{1+\alpha_0+A_0(\alpha_A-\alpha_B)}{q}\phi,
~~\psi\rightarrow\frac{\psi}{q},\nb\\
\alpha_A&\rightarrow&\frac{4\alpha(1+\beta)(1+\eta_0+A_0\eta_A)-\alpha_0}{A_0},~~~
\alpha_B\rightarrow\frac{(4\alpha(1+\beta)-\beta)(a+\eta_0+A_0\eta_A)-\eta_0-A_0\eta_A}{A_0},\nb\\
\omega&\rightarrow&\sqrt{1+\eta_0+A_0\eta_A},~~~m\rightarrow m\sqrt{(1+\beta)(1+\eta_0+A_0\eta_A)};
 \eqn

 \item transformations in the action Eq.(\ref{24b}) for p-wave superconductor:
 \bqn
 \lb{29}
\phi&\rightarrow&\frac{1+\xi_0+A_0\xi_A}{q}\phi, ~~~
\psi\rightarrow\frac{\psi}{q},~~~\eta_A\rightarrow\frac{A_0\xi_A+\xi_0-\eta_0-\beta-\beta\eta_0}{A_0(1+\beta)},\nb\\
\omega&\rightarrow&\sqrt{\frac{1+\xi_0+A_0\xi_A}{1+\beta}},~~~m\rightarrow m\sqrt{(1+\beta)(1+\xi_0+A_0\xi_A)};
 \eqn

\item transformations in the action Eq.(\ref{24c}) for d-wave superconductor:
\bqn
 \lb{30}
\phi&\rightarrow&\frac{1+\gamma_0-\gamma_AA_0\xi_A}{q}\phi, ~~~
\psi\rightarrow\frac{\psi}{q},~~~\gamma_0\rightarrow\eta_0+A_0\left(\eta_A-\gamma_A\right)+\beta(1+\eta_0+A_0\eta_A),\nb\\
\omega&\rightarrow&\sqrt{1+\eta_0+A_0\eta_A},~~
m\rightarrow m\sqrt{(1+\beta)\left(1+\eta_0+A_0\eta_A\right)}.
 \eqn
  \end{itemize}

After somewhat tedious but straightforward calculations, one obtains the following unified equation of motion in Ho\v{r}ava-Lifshitz gravity
\begin{subequations}\lb{31}
 \bqn \lb{31a}\psi''+\left(\frac{f'}{f}+\frac{2}{r}\right)\psi'+\left(\frac{\phi^2}{f^2}+\bar{V}_s(r)\right)\psi&=&0, \eqn
 \bqn \lb{31b}\phi''+\frac{2}{r}\phi'-\frac{2\psi^2}{f}\phi&=&0,\eqn
 \bqn \lb{31c}A_x''+\frac{f'}{f}A_x'+\left(\frac{\omega^2}{f^2}-2(1+\beta)\frac{\psi^2}{f}\right)A_x&=&0.\eqn
 \end{subequations}
where the effective potential is
 \bqn
 \lb{32}
 \bar{V}_s(r)=-\frac{m^2}{f}+(2-s)(s+(1-s)\alpha)\frac{f'}{rf}+\frac{\alpha}{2}(2-s)(1-s)\frac{f''}{f}+\frac{s(1-s)}{r^2}.
 \eqn
Here, similarly to the case in Einstein gravity, $s=0, 1, 2$ correspond to s-wave, p-wave and d-wave holographic superconductors, respectively.
The terms related to $\alpha$ and $\beta$ indicate the effects of the breaking Lorentz invariance in Ho\v{r}ava-Lifshitz gravity.
In other words, the holographic setup in Einstein gravity is restored with $\alpha=\beta=0$. Additionally, inspecting the Eq.(\ref{32}) one can see that only the s-wave state $(s=0)$ is sensitive to $\alpha$.

\begin{figure}[h]
\includegraphics[width=10cm]{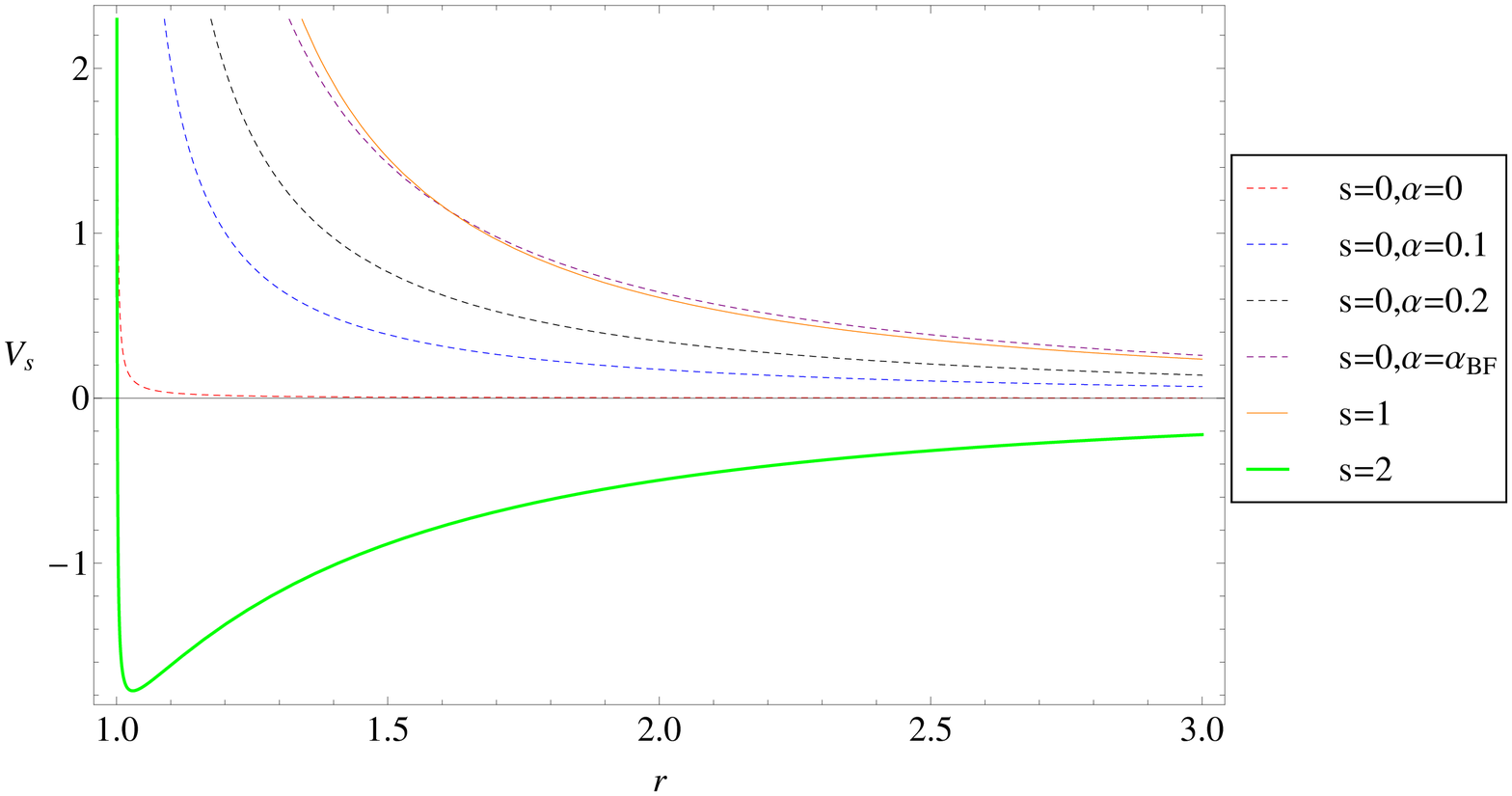}
\caption{Behaviour of $V_s$ for a black hole in HL gravity with $r_h=1,m^2_{BF}=-0.01,$ in the three different wave states. For this value of the squared mass $\alpha_{BF}=\frac{28}{75}$. } \label{fig00}
\end{figure}

\begin{figure}[h]
\includegraphics[width=10cm]{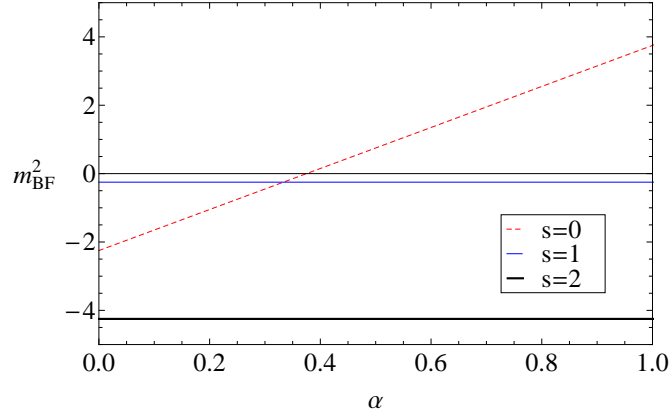}
\caption{ Breitenlohner-Freedman bound for s-wave $(m^2_{BF}=-\frac{9}{4}+6\alpha)$, p-wave $(m^2_{BF}=-\frac{1}{4})$ and d-wave $(m^2_{BF}=-\frac{17}{4})$ states. } \label{fig01}
\end{figure}

In Fig.(\ref{fig00}) one can observe that the behavior of the effective potential for the s-wave state with $\alpha_{BF}$ and p-wave state is quite similar. From the Eqs.($\ref{31a},\ref{32}$) the BF bound for the planar black hole in HL gravity can be obtained for each wave state and results in a more complex relation. It is
 \bqn
 \lb{17HL}
 m^2_{BF}=-\frac{9}{4} + 5 s - 3 s^2 + 6 \alpha - 9 s \alpha + 3 s^2 \alpha.
 \eqn
In general the BF bound is a negative value for the squared mass of the scalar field, however for the s-wave state this limit becomes positive at critical value $\alpha_c>\frac{3}{8}$ and increases linearly with $\alpha$. For d-wave and p-wave this bound is not affected by $\alpha$ as one can viewed in Fig.(\ref{fig01}).

\subsection{Numerical results on the condensation and the optical conductivity}

Now we are in position to solve the equations of motion Eqs.(\ref{31a}-\ref{31c}) and to numerically study the condensation and the conductivity of the superconductors.

Considering the background metric Eq.(\ref{22}), the boundary conditions of Eqs.(\ref{31a}-\ref{31c}) read
 \bqn
 \lb{33}
 \phi&=&\mu+\frac{\rho}{r},\nb\\
 \psi&=&\psi_1r^{-(3-\Delta)/2}+\psi_2r^{-(3+\Delta)/2},\nb\\
 A_x&=&a_0+\frac{a_1}{r},
 \eqn
where $\Delta$ satisfies
\bqn\lb{33b}
\Delta^2=9+4m^2-24\alpha-4s(5-3s+3s\alpha-9\alpha).
\eqn
According to gauge/gravity duality, $\mu$ and $\rho$ are interpreted as the chemical potential and the charge density of the theory on the boundary.
Either $\psi_1$ or $\psi_2$ can be the source of the dual operator, and therefore must take to be zero, while the other one is related to the vacuum expectation value of the dual operator $< O_i> \sim \psi_i(i=1, 2)$.
The optical conductivity can be read off from the asymptotic behavior of $A_x$ as
\bqn\lb{34}
\sigma(\omega)=\frac{a_1}{i\omega a_0}.
\eqn

Near the horizon, the regular condition implies that $\phi(r\to r_h)=0$ while $\psi(r\to r_h)=finite$, and the equation of motion gives $A_x(r\to r_h)\sim f^{\pm\frac{i\omega}{f'(r_h)}}$ among which we will choose the infalling solution with the minus sign.
The equations then can be numerically integrated from the horizon to the boundary via the shooting method.

We note one may rewrite Eq.(\ref{32}) by substituting $m$ in favor of $\Delta$ defined by Eq.(\ref{33b}) as follows,
\bqn
 \lb{32b}
 \bar{V}_s(r)=-\frac{(\Delta^2-9)r^3-4r_h^3s}{4r^2(r^3-r_h^3)}.
\eqn
Before numerically solving the equations, we give several remarks on the properties of the system.
Firstly, Eq.(\ref{32b}) indicates that the equation of motion can be view as {\it not} explicitly dependent on the value of $\alpha$, which measures the corrections of Ho\v{r}ava-Lifshitz theory. In the next section, however, we will see that the critical temperature depends on $\alpha$.

\begin{figure}[h]
\includegraphics[width=9cm]{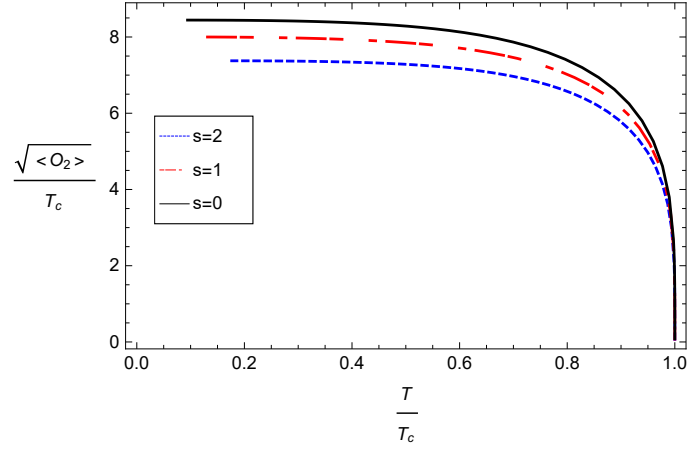}
\caption{Condensation curves $\psi\sim r^{-(3+\Delta)/2}$ for $\Delta=1$.} \label{fig1}
\end{figure}

Therefore, for a given $\alpha$ in Ho\v{r}ava-Lifshitz gravity, one can always find a corresponding system in Einstein gravity by tuning the value of $m$, so that the properties of condensate curve are the same.
However, since $\beta$ appears in Eq.(\ref{31c}), the conductivity of the two systems remains different.

\begin{figure}[h]
\includegraphics[width=8cm]{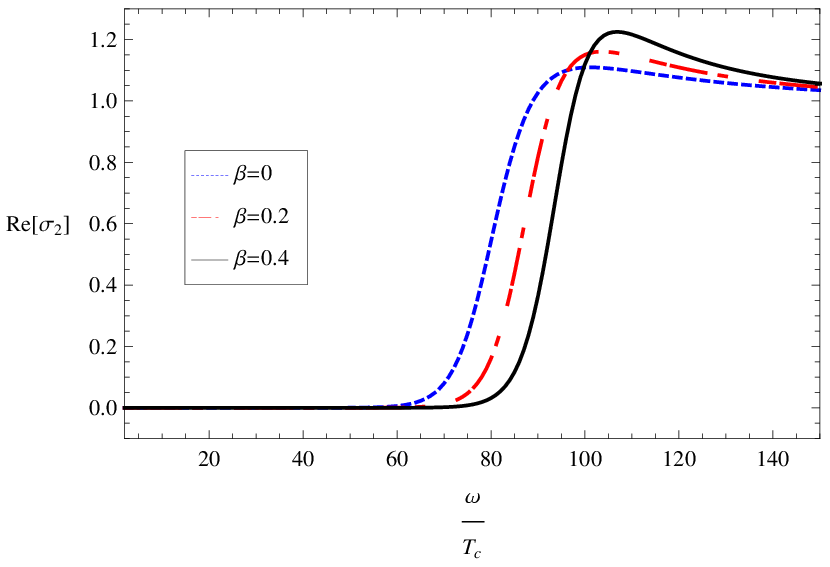}\includegraphics[width=8cm]{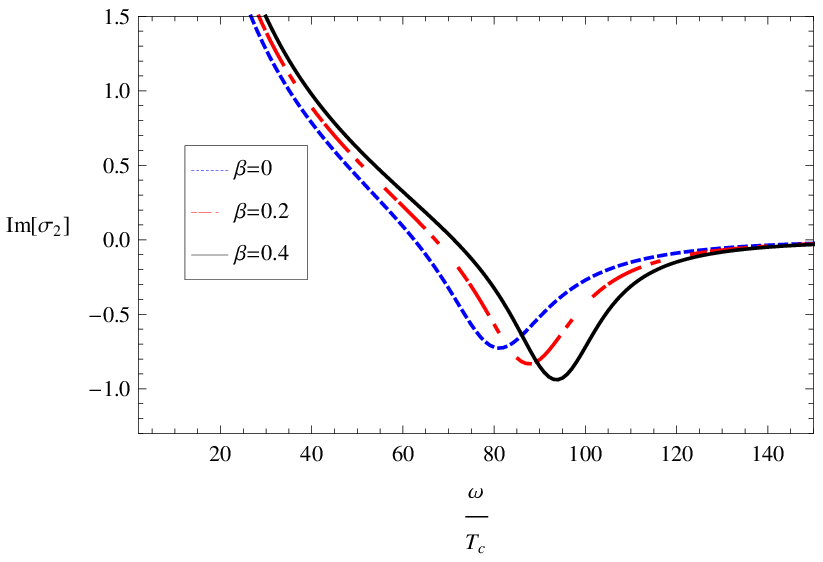}
\caption{Conductivity for $\Delta=1$, $s=0$, and $\psi\sim r^{-(3+\Delta)/2}$ at infinity.} \label{fig2}
\end{figure}

\begin{figure}[h]
\includegraphics[width=8cm]{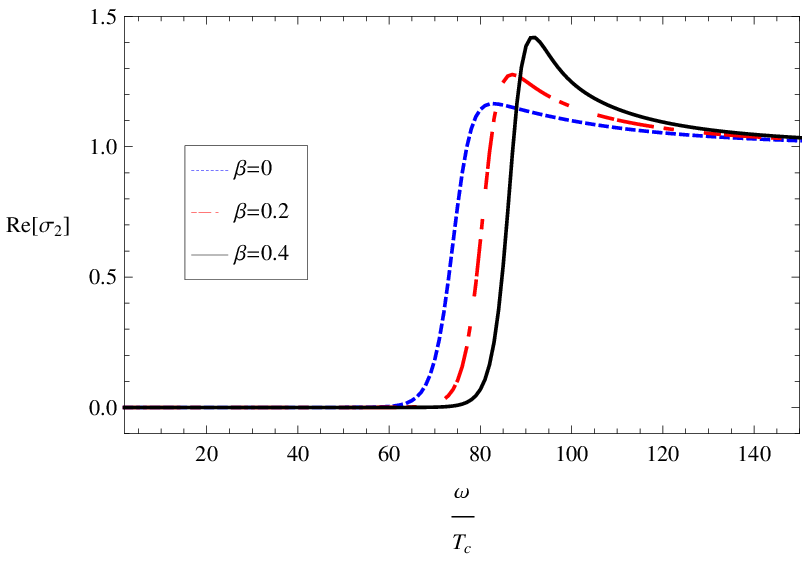}\includegraphics[width=8cm]{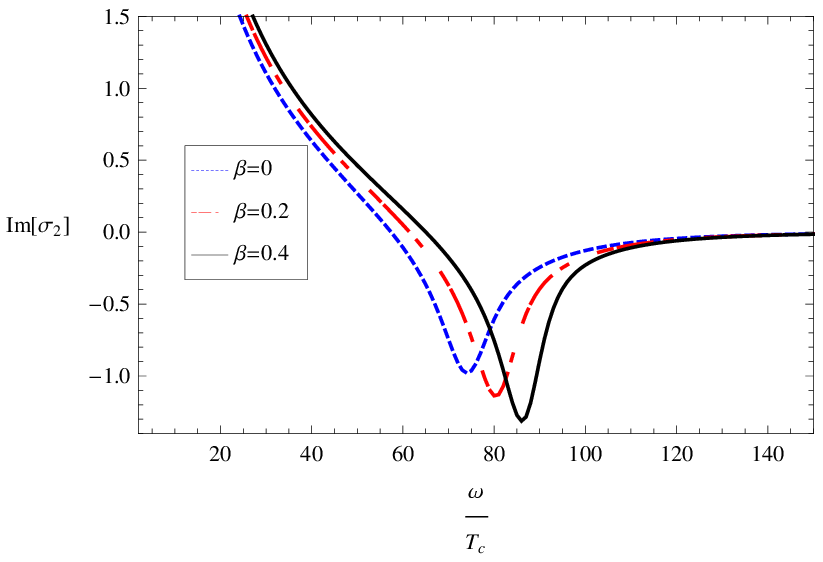}
\caption{Conductivity for $\Delta=1$, $s=1$ and $\psi\sim r^{-(3+\Delta)/2}$ at infinity.} \label{fig3}
\end{figure}

\begin{figure}[h]
\includegraphics[width=8cm]{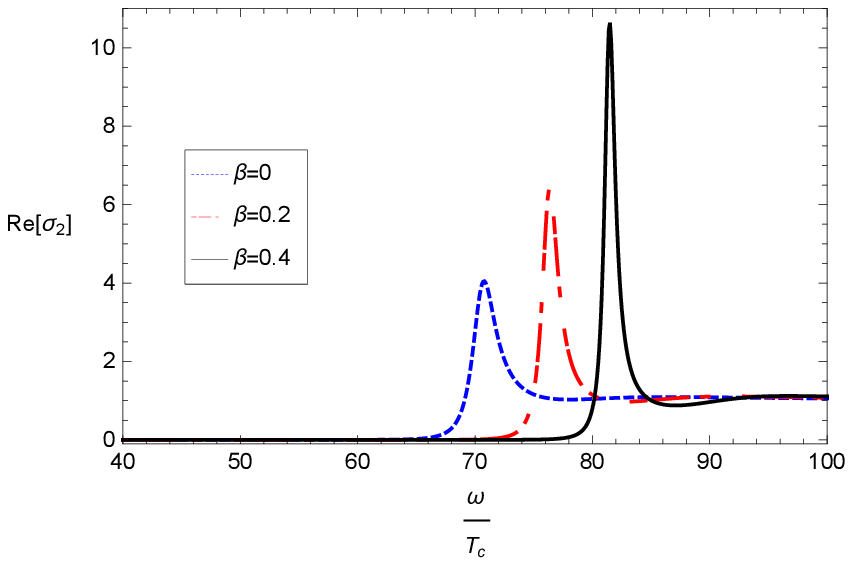}\includegraphics[width=8cm]{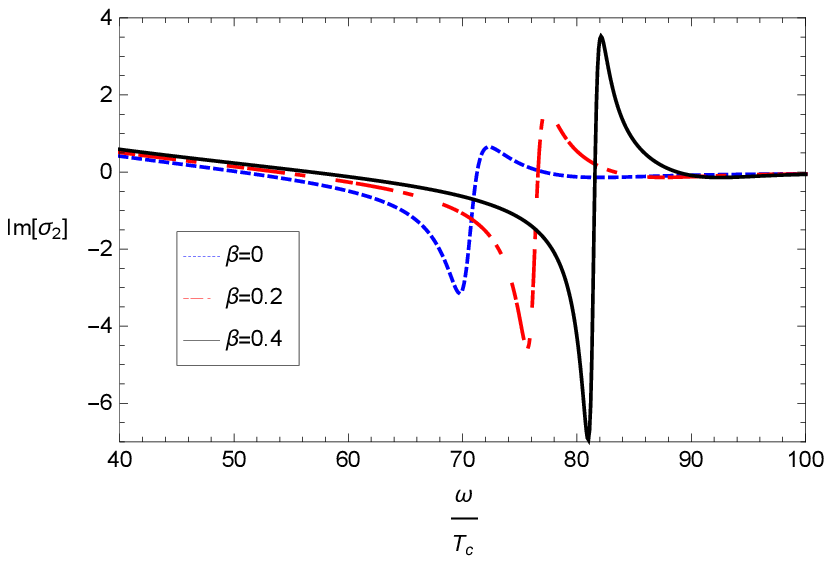}
\caption{Conductivity for $\Delta=1$, $s=2$ and $\psi\sim r^{-(3+\Delta)/2}$ at infinity.} \label{fig4}
\end{figure}

The numerical results for various holographic superconductors are in Figs.\ref{fig1}-\ref{fig4}.

In Fig.\ref{fig1}, we present the condensation of the matter field with $\Delta=1$ by solving Eqs.(\ref{31a}) and (\ref{31b}).
We study the condensate of $<O_2>$ by taking $\psi_1=0$ because as addressed in \cite{Hartnoll:2008kx}, the condensate of $<O_1>$ is unstable due to its tendency of divergence when the temperature approaches to zero.

Moreover, we show the results of optical conductivity $\sigma$ in Figs.\ref{fig2}-\ref{fig4}, which correspond to the behavior for s-wave, p-wave and d-wave superconducting phases, respectively.
From Eq.(\ref{31c}), one sees that the conductivity depends on the parameter $\beta$, which measures the breaking of the Lorentz invariance.
We observe that, as $\beta$ increases, the ratio between energy gap and the critical temperature also increases.
This result agrees well with the observation in the s-wave superconductor investigated in \cite{Lin:2014bya,Luo:2016ydt}.

%\begin{figure*}
%\includegraphics[width=6cm]{fig1_sig1A.eps}\includegraphics[width=6cm]{fig1_sig1B.eps}
%\caption{conductivity with $\Delta=1$ and $\psi\sim
%r^{-(3-\Delta)/2}$ at infinity, $s=0$ or $s=2$.} \label{fig2}
%\end{figure*}

\subsection{Semi-analytical results on the condensation}

In order to gain a better understanding of the influence of the HL gravity in holographic superconductor scenario, we have calculated a semi-analytical expression for the vacuum expectation value of the dual operator $< O_i>$ following the approach developed in \cite{Gregory:2009fj,Pan:2009xa}. Matching the asymptotic solution for $\phi(r)$ and $\psi(r)$ and their derivatives in an arbitrary point $z_m=\frac{r_h}{r_m}$ we get the following expression for the expectation value $<\mathcal{O}_i>$  as
\bqn
 \lb{O2}
 \left<\mathcal{O}_i\right>&=&\Pi~T^{\lambda_i}\left(\frac{T_c}{T}\right)\sqrt{1-\frac{T}{T_c}}\sqrt{1+\frac{T}{T_c}}
\eqn
where $\Pi$ is defined by
\bqn
 \Pi&=&\left(\frac{z_m^{-\lambda_i+1}\left(2-\sigma(1-z_m)\right)}{\lambda_i(1-z_m)+2 z_m}\right)\sqrt{\frac{3}{2(1-z_m)}}\left(\frac{4\pi}{3}\right)^{\lambda_i}\\
\eqn
and $\sigma$ and $\lambda_i$ as
\bqn
 \sigma&=&\alpha(2-s)(1-s)+s(2-s)-\frac{m^2}{3}\\
 \nonumber\\
 \lambda_i&=&\frac{3}{2}\pm\frac{\Delta}{2}
\eqn
The expression for the critical temperature $T_c$ is given by
\bqn
 \lb{OTc}
 T_c&=&\frac{3}{4\pi}\sqrt{\frac{\rho}{6}\sqrt{\frac{2}{\Sigma}}}\\
 \nonumber\\
 \Sigma&=&\frac{s}{2}+\frac{\sigma(\sigma-2)}{2}-\frac{\sigma}{1-z_m}+\frac{\lambda_i (2-\sigma(1-z_m))}{\lambda_i (1-z_m)+2z_m}\left(\frac{1}{1-z_m}\right)
\eqn
The results presented in the Eqs.(\ref{O2},\ref{OTc}) generalize that one in \cite{Gregory:2009fj} for $2+1$ holographic superconductor.
\begin{figure}
\includegraphics[width=8cm]{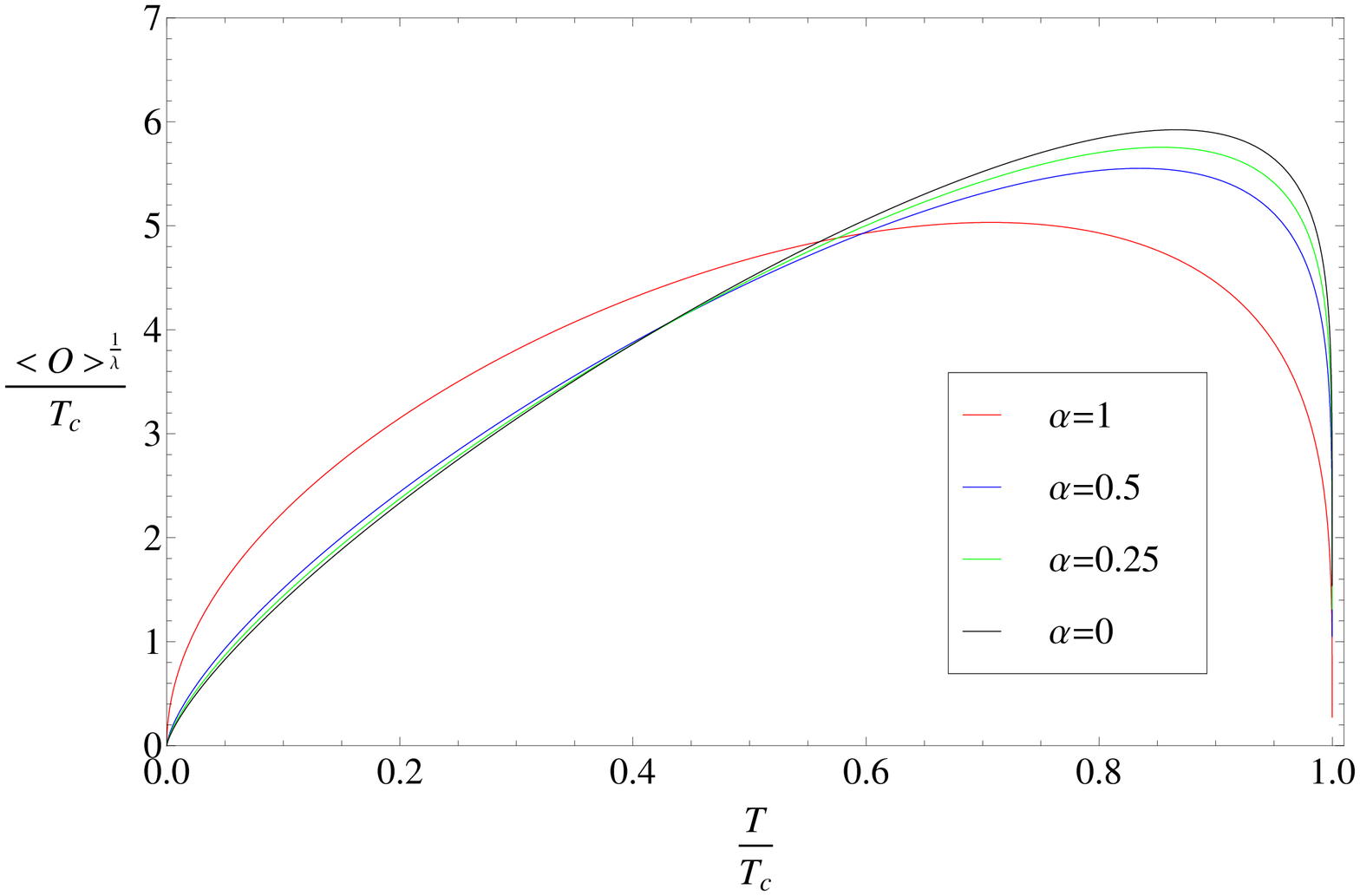}
\includegraphics[width=8cm]{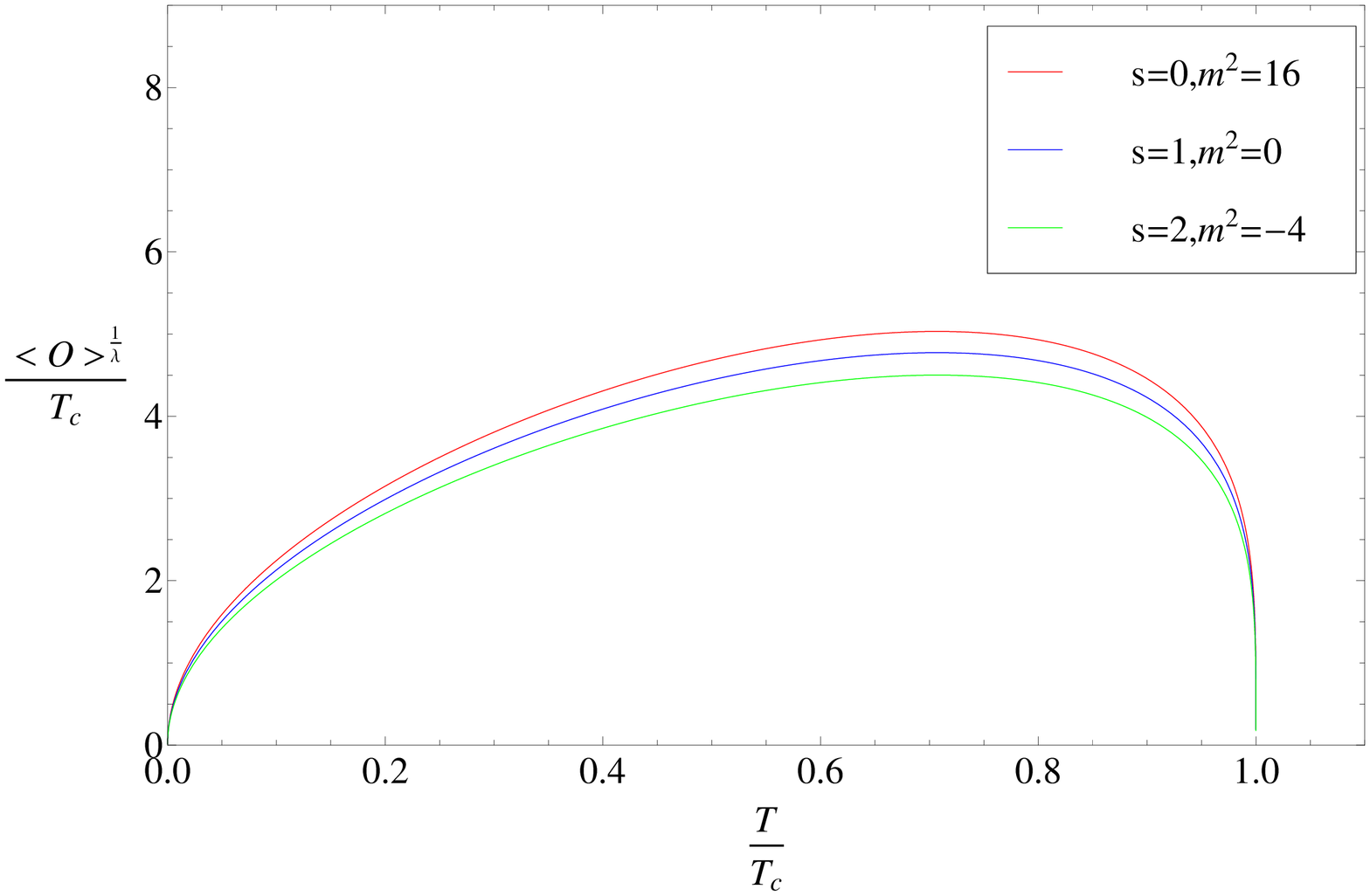}
\caption{Condensation curves obtained from the Eq.(\ref{O2}) for some values of $\alpha$ and $z_m=1/2$ for s-wave state superconductor (Left) and for the three wave states with $\alpha=3$ (Right).} \label{figtemp3}
\end{figure}
In Fig.(\ref{figtemp3}) one observes that the Eq.(\ref{O2}) fails in describing the behavior of the condensate when $T\to0$. However near to the critical temperature $T_c$ the mean field theory results $<\mathcal{O}_i>\sim\sqrt{1-\frac{T}{T_c}}$ is recovered. For the s-wave state, the condensation becomes easier when $\alpha$ increases. For the p-wave and d-wave states the condensation is not affected by $\alpha$. Thus, looking at only the phase transition curves the s-wave state is the only one that has a detectable influence of the HL gravity. Comparing the condensation among the three wave states one can see that a d-wave superconductor condensates at the highest temperature.

\begin{figure}
\includegraphics[width=8cm]{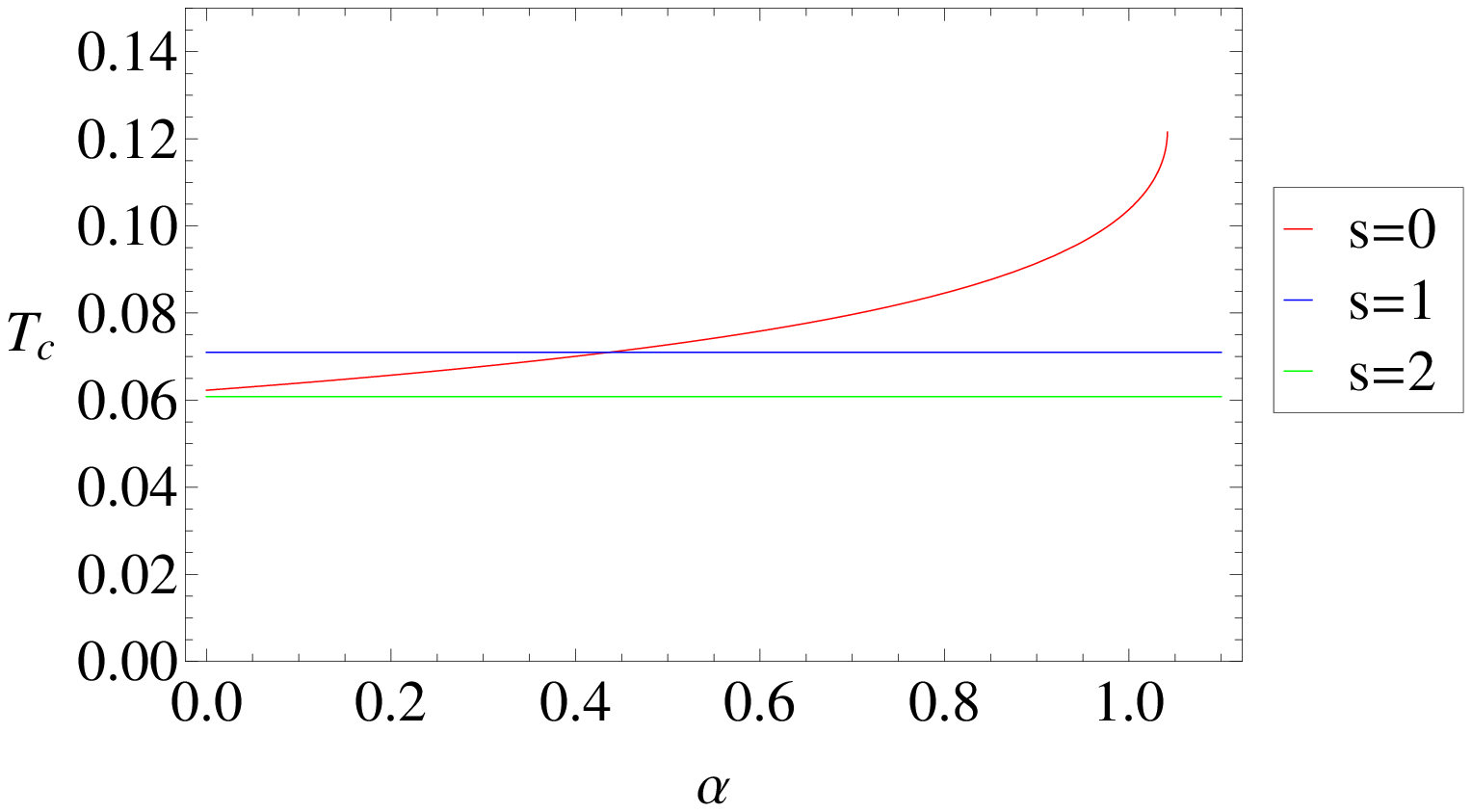}
\includegraphics[width=8.6cm]{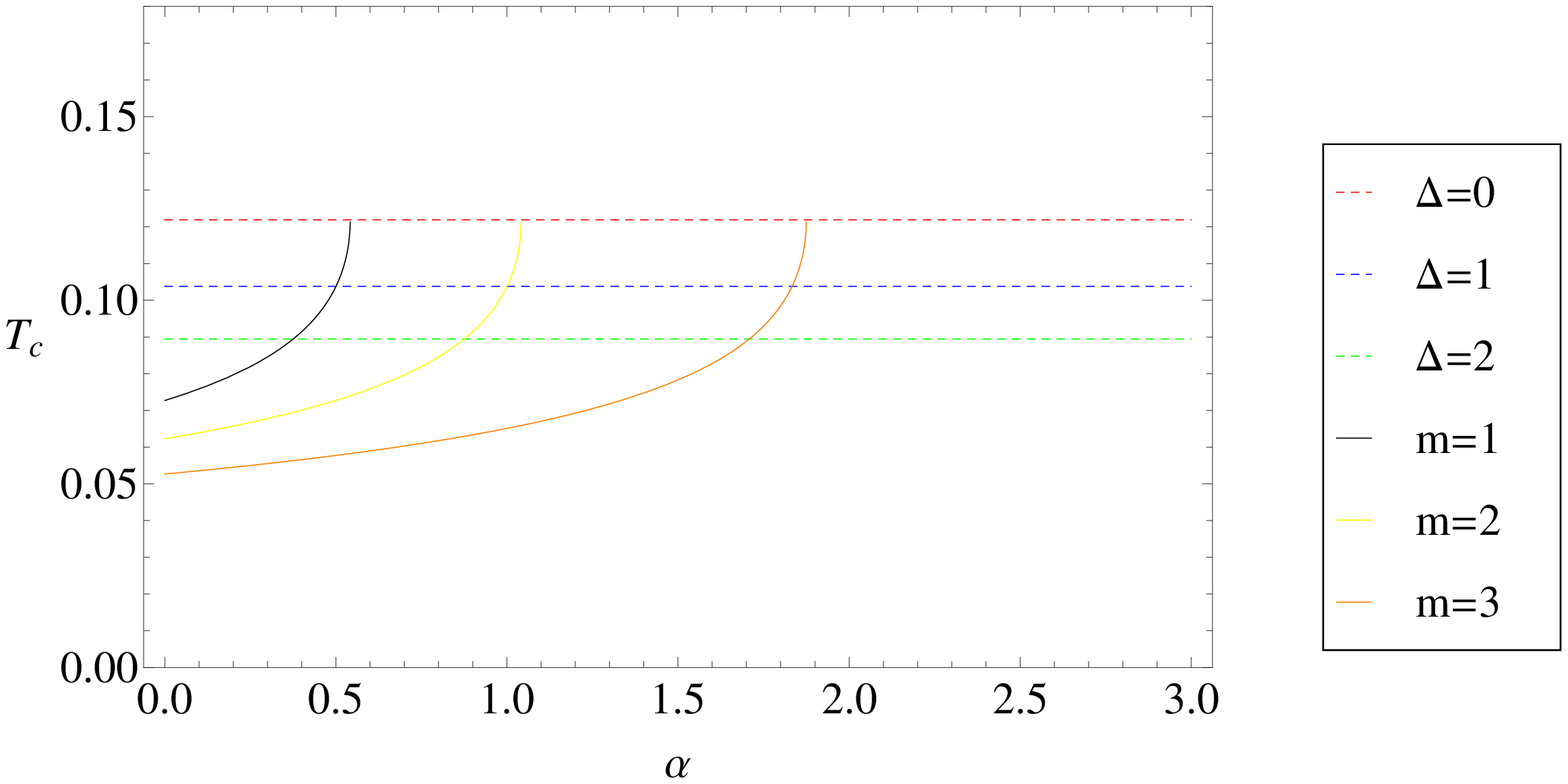}
\caption{Dependence of the critical temperature $T_c$ on $\alpha$ with $m=2$ for all three wave states (Left) and for some values of mass $m$ and $\Delta$ for the s-wave state (Right).}\label{figtemp}
\end{figure}

In Fig.(\ref{figtemp}) one can see that just as in BF bound, the critical temperature is independent of $\alpha$ for the d-wave and p-wave states. In general, $T_c$ has an upper bound with $\alpha$ for the s-wave state. The horizontal lines show the values of $T_c$ when $\Delta=0,1,2$.

\section{Concluding remarks}
\renewcommand{\theequation}{4.\arabic{equation}} \setcounter{equation}{0}

In summary, in this work, we study the s-wave, p-wave, and d-wave holographic superconductors dual to a planar black hole in Ho\v{r}ava-Lifshitz gravity.
For the models discussed in the present work, we manage to rewrite the equations of motion into an unified form.
Numerical  and semi-analytical calculations are carried out and condensations are found for all three holographic superconductors with different wave states. The s-wave superconductors suffer the strongest influence of HL gravity parameters. Thus, they could be used to test the Lorentz symmetry breaking.
However, the present study is limited to the case of planar AdS background, a more general metric may likely lead to more divergent results.
We did neither consider the backreaction of the matter field, which also may imply additional nontrivial effects owing to the breaking of Lorentz invariance.
Lastly, we have only discussed the lower energy sector.
As a result, many effects such as superluminal particle and universal horizon lay beyond the scope of the current study.
Nonetheless, the calculated results show that the parameter $\beta$ plays a distinct role in the properties of the optical conductivity even for a rather simple metric.
Ho\v{r}ava-Lifshitz gravity is one of the theories that successfully passes all the observable post-Newtonian test via solar data.
In this context, the findings of the present work somewhat indicate that the measured optical conductivity can be utilized to quantify the breaking of Lorentz invariance, and therefore shed some light on further discrimination between different theories.

\section*{\bf Acknowledgements}

We gratefully acknowledge the financial support from Brazilian funding agencies
Funda\c{c}\~ao de Amparo \`a Pesquisa do Estado de S\~ao Paulo (FAPESP),
Conselho Nacional de Desenvolvimento Cient\'{\i}fico e Tecnol\'ogico (CNPq),
and Coordena\c{c}\~ao de Aperfei\c{c}oamento de Pessoal de N\'ivel Superior (CAPES);
from Chinese funding agenicies
Natural Science Foundation under Grant No. 11705161, No. 11573022,  No. 11375279, No. 11775076 and No. 11690034,
Natural Science Foundation of Jiangsu Province under Grant No. BK20170481;
and from Chilean funding agency FONDECYT under grant No.3150006.

%%%%%%%%%%%%%%%%%%%%%%%%%%%%%%%%%%%%%%%%%%%%%%%%%%%%%%
%\bibliographystyle{../preprint}
%\bibliography{../Bib/QuantGra.bib}

\end{document}